\documentclass[prl,aps,twocolumn,superscriptaddress]{revtex4}
\usepackage{latexsym}
\usepackage{graphicx}

\newcommand{\be}{\begin{equation}}
\newcommand{\ee}{\end{equation}}
\newcommand{\ba}{\begin{eqnarray}}
\newcommand{\ea}{\end{eqnarray}}
\newcommand{\baa}{\begin{eqnarray*}}
\newcommand{\eaa}{\end{eqnarray*}}

\def\be{\begin{equation}}
\def\ee{\end{equation}}
\def\ba{\begin{eqnarray}}
\def\ea{\end{eqnarray}}

\def\C60{A$_x$C$_{60}$}

\def\HgCu3{HgCa$_2$Cu$_3$O$_{8+y}$}
\def\HgCu4{HgBa$_2$Ca$_3$Cu$_4$O$_{10+y}$}
\def\TlCu{Tl$_2$Ba$_2$CuO$_{6+\delta}$}
\def\TlCu3{Tl$_2$Ba$_2$Ca$_2$Cu$_3$O$_{10+y}$}
\def\TlCu4{Tl$_2$Ba$_2$Ca$_3$Cu$_4$O$_{12+y}$}

\def\BiCu3{Bi$_2$Sr$_2$Ca$_{2}$Cu$_3$O$_y$}

\def\8LSCO{La$_{1.88}$Sr$_{.12}$CuO$_4$}
\def\110LNSCO{La$_{1.5}$Nd$_{0.4}$Sr$_{0.1}$CuO$_{4}$}
\def\stage4LCO{La$_{2}$CuO$_{4+\delta}$}
\def\Y248{YBa$_2$Cu$_4$O$_8$}

\def\NbSe2{NbSe$_2$}
\def\TaSe2{TaSe$_2$}
\def\TiSe2{TiSe$_2$}

\begin{document}
\title{S-wave superconductivity with orbital dependent sign change in the checkerboard models of iron-based superconductors}
\author{Xiaoli Lu }
\affiliation{Center for Statistical and Theoretical Condensed
Matter Physics, and Department of Physics, Zhejiang Normal
University, Jinhua 321004, China}
\affiliation{Department of Physics, Purdue University, West
Lafayette, Indiana 47907, USA}
\author{Chen Fang}
\affiliation{Department of Physics, Purdue University, West
Lafayette, Indiana 47907, USA}
\author{Wei-Feng Tsai}
\affiliation{Institute for Solid State Pysics, University of Tokyo, Kashiwa 277-8581, Japan}
\author{Yongjin Jiang}
\affiliation{Center for Statistical and Theoretical Condensed
Matter Physics, and Department of Physics, Zhejiang Normal
University, Jinhua 321004, China} \affiliation{Department of Physics, Purdue University, West
Lafayette, Indiana 47907, USA}
\author{Jiangping Hu}
 \altaffiliation{Corresponding author}
\email{hu4@purdue.edu} \affiliation{Department of Physics, Purdue
University, West Lafayette, Indiana 47907, USA} \affiliation{Beijing
National Laboratory for Condensed Matter Physics, Institute of
Physics, Chinese Academy of Sciences, Beijing 100080, China}

\begin{abstract}
We study three different 
 multi-orbital models for iron-based superconductors (iron-SCs) in the solvable limit of weakly 
 coupled square plaquettes. 
The strongest superconducting  (SC) pairing is in the $A_{1g}$ $s$-wave
channel and its development 
is correlated with the emergence of the next-nearest-neighbour
antiferromagnetism (NNN-AFM). For the models with more than three
orbitals, this study suggests that the signs of the intra-orbital
pairing order parameters of the $d_{xy}$ and the $d_{xz}$ (or
$d_{yz}$) orbitals must be {\it opposite}. 
Such sign difference stems from
the intrinsic symmetry properties of inter-orbital hoppings and 
might, ultimately, lead to the sign-change of the SC orders between the 
hole Fermi pockets at the $\Gamma$ point and produce anisotropic or even
gapless SC gaps in the electron  Fermi pockets around the $M$ point in reciprocal space,  as restoring back to the homogeneous limit.
\end{abstract}
\maketitle

{\it Introduction.}- Since the 
 iron-based superconductors were discovered two years ago\cite{kamihara-08jacs3296}, the relation between the new
superconductors and the high-$T_c$ cuprates has been a central focus of
researches. It is highly debated that whether the new superconductors belong to
the same category of strongly correlated electron systems in which the cuprates are believed to be.
  Models based on both  strong coupling
\cite{seo-08prl206404,si-08prl076401,Berg2010,Fang2008,Xu2008a,dai2008j,Haule2009,Haule2008}
and weak coupling\cite{mazin-08prl057003,kuroki-08prl087004,wang-09prl047005,
thomaleasvsp,cvetkovic-09epl37002,korshunov-08prb140509,chubukov-08prb134512}
approaches have been applied to understand the properties of the new materials and their
relation to the curpates.

Strong correlation in an electron system can be reflected in its
strong ``locality'' of physical properties caused by 
 short-range interactions. For instance, both magnetism and superconductivity in the cuprates
  could be attributed to such ``local'' physics.
The magnetism in the parent cuprate compounds
is well described 
by the Heisenberg model with the nearest-neighbor (NN) antiferromagnetic (AFM) exchange
couplings.
The $d$-wave superconductivity  with symmetry form factor, $\cos k_x-\cos k_y$, in reciprocal space\cite{scalapino1995}
 corresponds to short-range superconducting (SC) pairings between just the NN copper sites.
Theoretically, the locality (short coherence length) already allows us to understand many
 essential physics from small clusters and to give new insights for obtaining low-energy
  effective models\cite{scalapino1996,tsai2006,yaoh2007,altman2002}.

Remarkably, recent experimental results in iron-based superconductors suggest
that both magnetism and superconductivity display an excellent
locality behavior as well. Neutron scattering experiments
demonstrate that the magnetism in the parent compounds of iron-SCs can be  again described 
well by the Heisenberg 
model with the NN 
and the NNN-antiferromagnetic exchange couplings between iron spins\cite{Cruz2008,Zhaoj2009,Zhaojun2008}.  In addition, angle resolved photo emission spectroscopy (ARPES) experiments show that the SC gaps can be
fitted to a single $\cos k_x\cos k_y$ functional form in reciprocal
space\cite{Ding2008a,Zhang2010d,Nakayama2010}. 
Regarding the fact that iron-SCs are complicated multi-orbital systems,  these experimental results are
 compelling evidence supporting that a local interaction  controls the pairing channels
since the function form in the real space 
 simply corresponds to SC pairings between the NNN iron sites. They motivate us to ask whether the physics of iron-SCs can also be understood
from 
checkerboard-like models as studied in Ref. \cite{tsai2006}, where the local physics (within a plaquette)
can be solved exactly, with an assumption that the major physics would remain unchanged
as the models are adiabatically tuned to the homogeneous limit.

\begin{figure}[tbh]
\includegraphics[width=6cm, height=5cm]{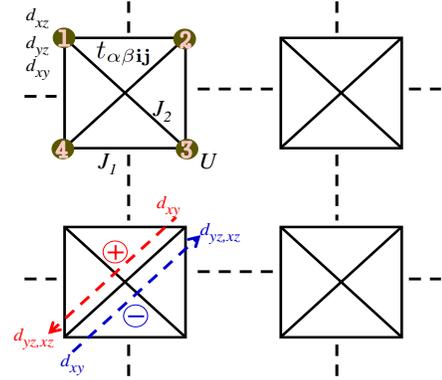}
\caption{The sketch of the checkerboard models. The hopping signs between $d_{xy}$ and $d_{xz(yz)}$ orbitals are indicated. \label{checkerboard}}
\end{figure}

In this Letter, we present results for three different 
tight-binding models for iron-SCs which are constructed using
two \cite{raghu2008}, three \cite{daghofer2010}, and four
orbitals\cite{Daghofer2008}, respectively, on a checkerboard lattice (see Fig.~\ref{checkerboard}).  Based on the knowledge of exactly solvable four-site problems,
we show the phase diagrams of the models in the limit of weakly coupled plaquettes as a function of intra-orbital onsite interaction $U$, inter-orbital onsite interaction $U^\prime$, 
Hund's coupling and onsite inter-orbital pairing hopping $J_H$, as well as the NNN-AFM coupling $J_2$ which can be generated by superexchange mechanism mediated by As atoms. In all of the models, the leading SC phase is always from the $A_{1g}$ $s$-wave pairing channel in reasonable parameter regions. The superconductivity is
intimately correlated with the development of the NNN-AFM and
becomes stronger as $J_2$ increases. Most remarkably, this
study also shows that the superconductivity and magnetism are
orbital-selective in the three- and four-orbital models: First, in
the three-orbital model, the AFM is more pronounced in the $d_{xy}$
orbital while the superconductivity is more pronounced in the $d_{xz}$
and $d_{yz}$ orbitals. Second, in both three- and four-orbital
models, the signs of the intra-orbital pairing order parameters of
the $d_{xy}$ and the $d_{xz}$ (or $d_{yz}$) orbitals are opposite.
The sign difference stems from the intrinsic symmetry of the
inter-orbital hopping amplitudes between $d_{xy}$ and
$d_{xz(yz)}$\cite{Lee2008a}. 
This feature could result in the
sign-change of the SC order parameters between the hole Fermi pockets at the
$\Gamma$ point and produce anisotropic or gapless SC gaps in the electron pockets around $M$ point in reciprocal space  in the homogeneous limit.

{\it Model and method.}- A generic Hamiltonian of iron-SCs can be written as $H=H_{0}+H_{I}$, where $H_{0}$ is the kinetic energy of the $d$-electrons of irons and $H_I$ describes the interactions between them.
 Explicitly, $H_0$ is given by
\begin{equation}
H_0= \sum_{\bf i,j,\sigma}\sum_{\alpha,\beta} (t^{\bf ij}_{\alpha\beta}+\epsilon_\alpha\delta_{\alpha\beta})
d^{\dagger}_{{\bf i},\alpha,\sigma} d^{\phantom{\dagger}}_{{\bf
j},\beta,\sigma}+H.c.
\end{equation}
where $\alpha,\beta$ are orbital indices, ${\bf i},{\bf j}$ label sites and
$\sigma$ is spin index. $H_I$ can be written as $H_I=
\sum_iH_{Io}(i)+H_{Ie}$, where $H_{Io}(i)$ is the onsite interaction,
given by,
\ba
H_{Io}({\bf i}) &=& U\sum_{\alpha} n_{{\bf i},\alpha,\uparrow}n_{{\bf
i},\alpha,\downarrow} + \sum_{\alpha\neq \beta}[\frac{U^\prime}{2} n_{{\bf
i},\alpha}n_{{\bf i},\beta} - \frac{J_{H}}{2} S_{i\alpha}S_{i\beta}] \nonumber \\
&+& J\sum_{\alpha\neq\beta}d^\dagger_{{\bf i},\alpha,\uparrow}
d^\dagger_{{\bf i},\alpha,\downarrow}
d_{{\bf i},\beta,\downarrow}d_{{\bf i},\beta,\uparrow},
\ea
 with $U$, $U^\prime$, $J_H$, and $J (=J_H)$ denoting intra-orbital repulsion, inter-orbital repulsion, ferromagnetic Hund's coupling, and inter-orbital pair hopping, respectively.
$H_{Ie}$ describes interactions between different sites. In this
paper, we specifically consider the NNN (denoted by double angle brackets) magnetic exchange coupling $J_{2}$, which can be naturally generated by the superexchange mechanism through As atoms,
\be
H_{Ie}=\sum_{\langle\langle{\bf i,j}\rangle\rangle}\sum_{\alpha,\beta} J_{2}{\bf S}_{{\bf
i},\alpha}\cdot {\bf S}_{{\bf j},\beta}.
\ee

\begin{figure}[tb]
\includegraphics[width=7cm]{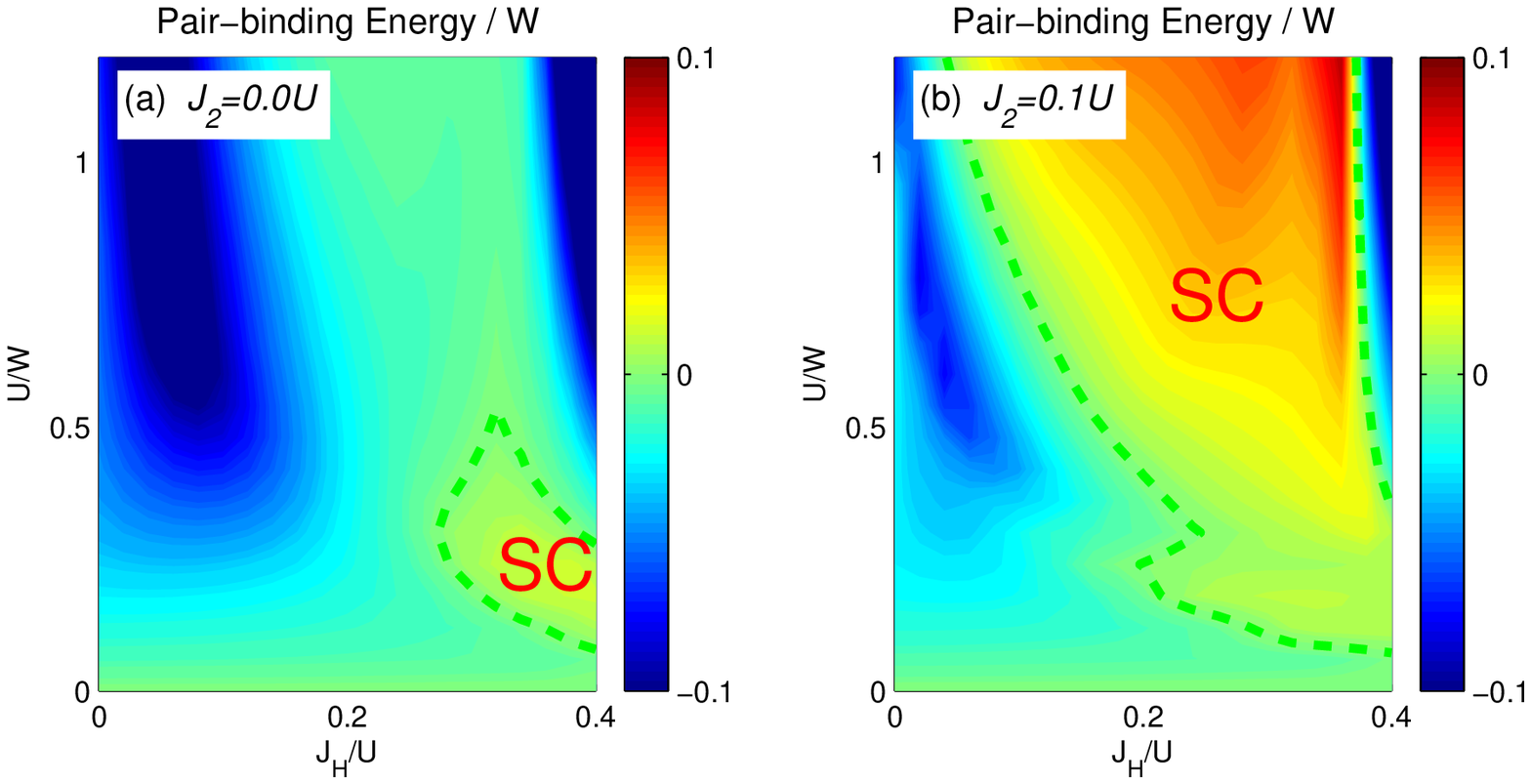}
\includegraphics[width=7cm]{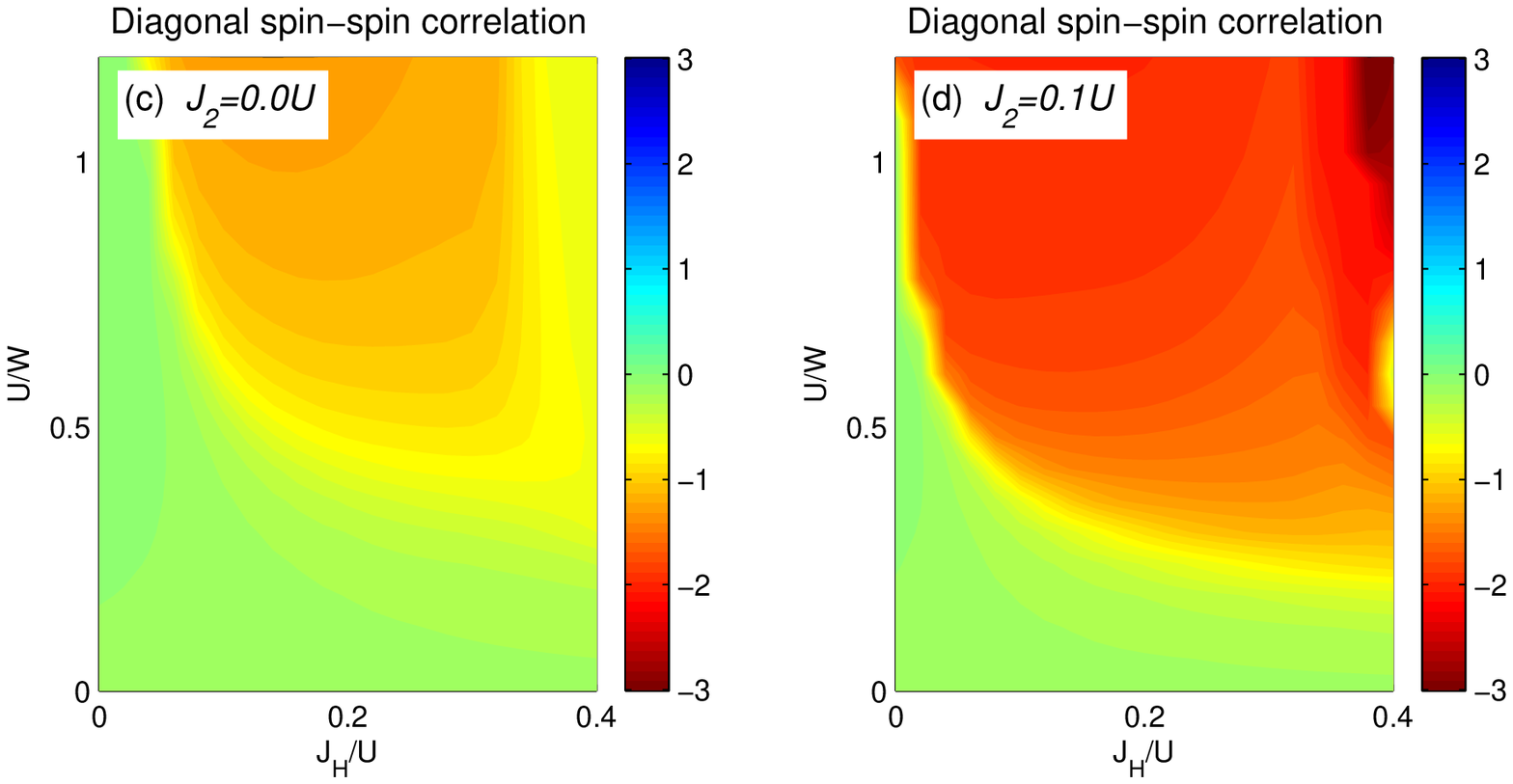}
\caption{The pair-binding energy [(a) and (b)] and the diagonal spin-spin
correlations [(c) and (d)] for $2\times 2$ plaquette in the
three-orbital model 
are plotted as functions of $U/W$ and $J_H/U$ with [(a) and (c)] and without
[(b) and (d)] exchange coupling $J_2$. The dotted lines in
(a) and (b) enclose the regions with positive pair-binding energy,  suggesting a SC phase. \label{pairing_energy}}
\end{figure}

Different tight-binding models have been proposed to describe the
band structures of iron-SCs. In this study, we take
three different models: a two-orbital model given
in \cite{raghu2008}, a three-orbital model given
in \cite{daghofer2010}, and a four-orbital model in \cite{Daghofer2008}.
The tight-binding parameters of the models can be found in the above
references. We have also tried different four-orbital models reduced
from the five-orbital models constructed in\cite{kuroki-08prl087004}. All of the models capture the basic band structures of iron-SCs. In fact, the major results reported 
below are consistent in all of these models.


The checkerboard models are defined on the lattice shown in Fig.~\ref{checkerboard}, with the inter-plaquette hopping amplitudes $\tau^{ij}_{\alpha\beta}$ and exchange couplings $\mathcal{J}_2$ less than intra-plaquette parameters, $t^{ij}_{\alpha\beta}$ and $J_2$. A four-site plaquette can be diagonalized exactly, though one may necessarily resort to the numerics due to its multi-orbital complexity. Consequently, we can determine the pair-binding energy $E_p=2E(1)-E(0)-E(2)$, where $E(Q)$ is the ground-state energy for a given $Q$, the number of doped holes in an ``undoped'' reference state. In particular, $E_{p}>0$ indicates an effective attraction between doped holes. At generic doping, we consider the case $0\leq\{\tau_{\alpha\beta}^{ij},\mathcal{J}_2\}\ll E_p\ll \{t_{\alpha\beta}^{ij},J_2\}$, which allows us to obtain a controlled perturbation expansion of the full Hamiltonian by small inter-plaquette parameters. Second order perturbation gives us an effective theory describing interacting bosons ($Q=2$ states) with hopping integrals and short-range density-density interactions of order $\tau^2/E_p$ on the effective lattice. Furthermore, it can be proved that the ground state enters superfluidity at $T=0$\cite{Fisher1988}, except at special doping percentages. Thus, the positiveness of $E_p$ could be used as an economic way to identify SC phase with generic doping.

\begin{figure}[bt]
\includegraphics[width=7cm]{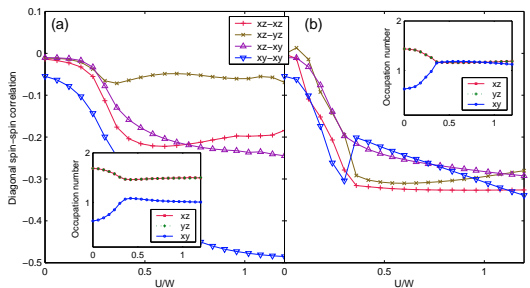}
\caption{The orbital-resolved diagonal spin-spin correlations,
$\langle\textbf{S}^1_{\alpha}\cdot\textbf{S}^3_{\beta}\rangle$
($\alpha,\beta=xz,yz,xy$), of a $2\times 2$ cluster with 16 electrons (2/3 filling)
(a) and 14 electrons (b). The parameters are given by $J_H=0.2U$, $J_2=0.1U$. 
Insets: The occupation number
of different orbitals as a function of $U/W$ with 16 electrons and 14
electrons, respectively.\label{spin_correlation}}
\end{figure}

{\it Spin and pairing correlations.}- The two-orbital model consists of the $d_{xz}$ and $d_{yz}$ orbitals and has been studied both analytically and numerically due to its simplicity.  However,
there is no SC phase in the physically meaningful parameter region in our study
if $J_2=0$. This  sharp feature is very similar to the results obtained from the functional renormalization group (fRG) for such a model\cite{thomaleasvsp}, although the
fRG is, in principle, a weak-coupling approach. When $J_2$ is added,
the superconductivity and magnetism in these two orbitals
 behave no qualitative difference from those in three- and four-orbital models within the same orbitals.
Therefore, we will focus on reporting the results from the three- and four-orbital models hereafter.

We summarize the main results for the three-orbital model in
Figs.~\ref{pairing_energy}, \ref{spin_correlation}, and \ref{pairing
amplitude}.  Note that the parent compounds here refer to 2/3 filling, instead of 1/2 filling in two- and four-orbital models. In Fig.~\ref{pairing_energy}, we plot the pair-binding energy
and the diagonal (NNN) spin-spin correlation function as a function of
$U/W$ and $J_H/U$ with $U'$ satisfying the $SU(2)$ symmetry
condition $U'=U-2.5J_H$ and $W$ being the bandwidth. As shown in the
figure, without $J_2$, there is a small region (circled by the dash
line) where the SC state is favored and this region is located within the region the NNN-AFM is developed. With a finite $J_2=0.1U$, the SC region is significantly enlarged as well as the NNN-AFM correlation. It is also important to note that a finite $J_H$ is needed in order to achieve positive pair-binding energy.

 The NNN spin-spin correlation within a plaquette, 
 $\langle Q;\text{GS}|\textbf{S}^{1}_{\alpha}\cdot\textbf{S}^{3}_{\beta}|Q;\text{GS}\rangle$, as a function of $U/W$ is shown in Fig.~\ref{spin_correlation} with fixed $J_H=0.2U$ and $J_2=0.1U$. The NNN-AFM correlation develops when $U$ reaches close to $0.4W$ and
the intra-orbital correlations are much stronger than the inter-orbital correlations. Moreover, the $d_{xy}$ orbital has the strongest NNN-AFM correlation as expected since the $d_{xy}$ orbital is half filled  [see Inset of Fig.~\ref{spin_correlation}(a)]. After doping two holes per plaquette, the correlations
become even among all the orbitals
,   as suggested by equally distributed occupation number among all orbitals [see Inset of Fig.~\ref{spin_correlation}(b)].

\begin{figure}
\includegraphics[width=7cm]{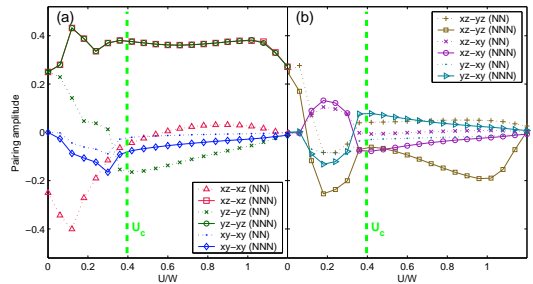}
\caption{The pairing amplitude of the intra-orbital (\textbf{a})
and the inter-orbital (\textbf{b}) electron pairs as function of $U$
at fixed parameters of $J_H=0.2U$ and $J_2=0.1U$. The green
dotted lines divide the regions of positive (right) and negative
(left) pairing energy. The `NN' in the legend means sites 1\&2 while
`NNN' means sites 1\&3; all other pairings' amplitude can be derived
from the $A_{1g}$ symmetry.\label{pairing amplitude}}
\end{figure}

The SC pairing amplitudes, defined as $\langle Q=2;\text{GS}|\Delta_{\alpha\beta}(\textbf{ij})|Q=0;\text{GS}\rangle$
with $\Delta_{\alpha\beta}(\textbf{ij})=d_{i,\alpha,\downarrow}d_{j,\beta,\uparrow}
-d_{i,\alpha,\uparrow}d_{j,\beta,\downarrow}$, are shown in Fig.~\ref{pairing amplitude}.
 The positive pair-binding energy is achieved when $U>U_c\sim 0.4W$, close to where the NNN-AFM has
 well developed.  As $E_p>0$, it is important to notice that due to $C_{4v}$ symmetry of a single plaquette and the
 fact $|Q=0;\text{GS}\rangle,|Q=2;\text{GS}\rangle$ belonging to $A_{1g}$ representation, the pair-field operator must
  have precisely $A_{1g}$ $s$-wave symmetry. The intra-orbital NNN pairings in the $d_{xz(yz)}$ orbitals dominate the pairing strength. The pairing in the $d_{xy}$ orbital is small since the $d_{xy}$ orbital remains 
half-filled and its AFM correlation is too strong. However, this is an important feature that even as the pairing strength is small in the $d_{xy}$ orbital, the sign of the pairing is opposite to the ones in the $d_{xz(yz)}$ orbitals. We find that the sign-change is as universal as the $A_{1g}$ s-wave pairing symmetry in all of the parameter regions we calculate.

\begin{figure}
\includegraphics[width=4cm]{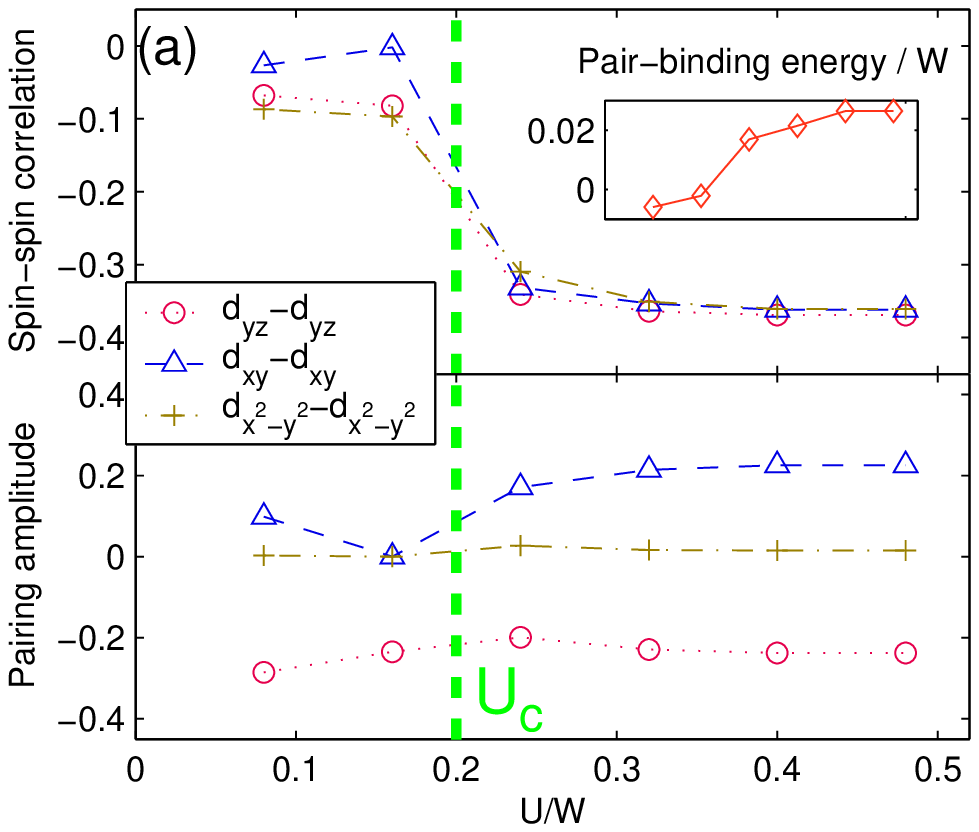}
\includegraphics[width=3.5cm]{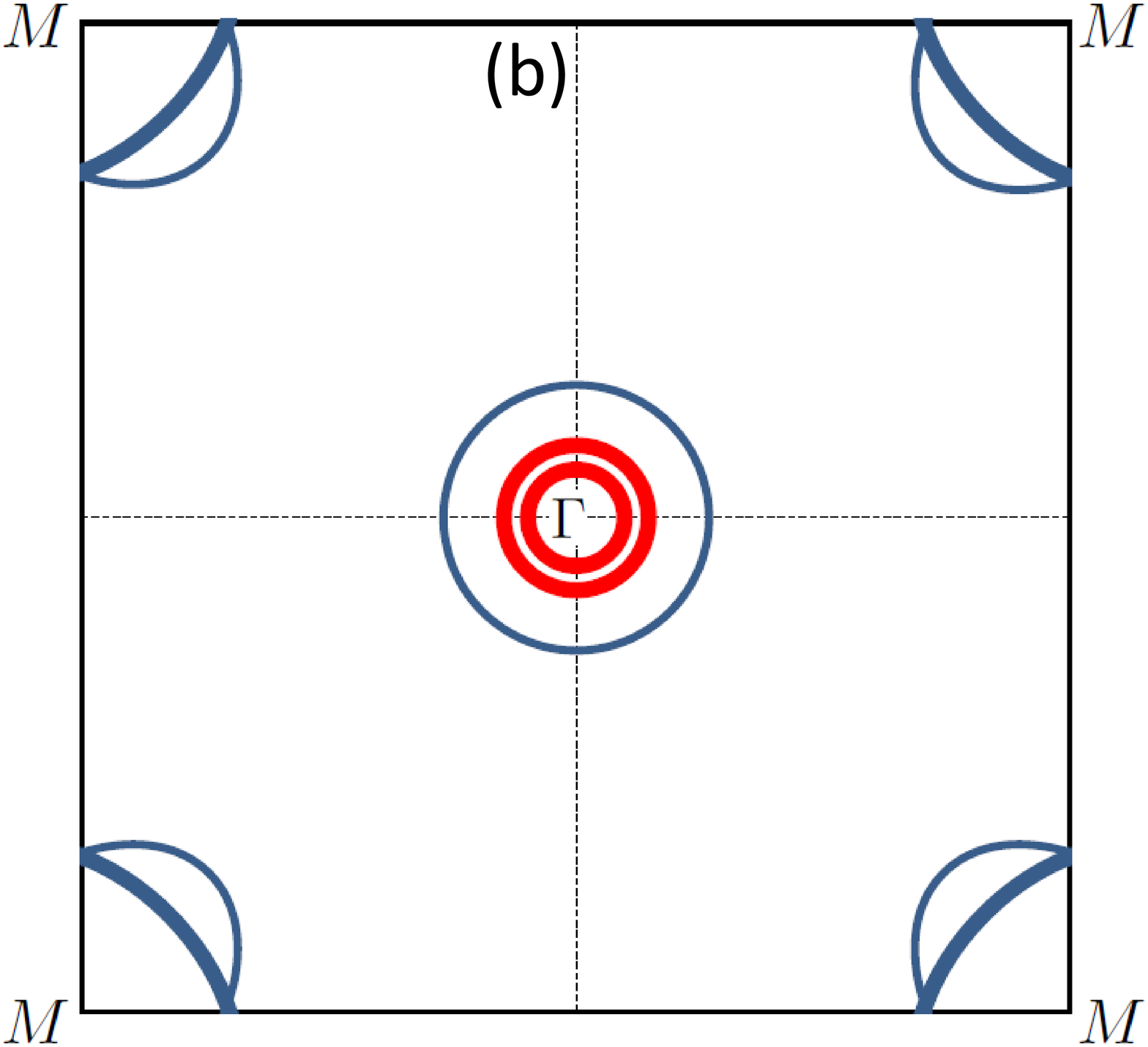}
\caption{\textbf{a}: The spin correlation (upper) and pairing amplitude (lower) of diagonal bonds as functions of $U$ in the four-orbital model. The inset shows the pair binding energy in unit of band width. \textbf{b}: The schematic pairing amplitude of a 2D five-orbital model in the superconducting state inside the first (folded) Brillouin zone. Red and blue means positive and negative signs, and the thickness indicates the size of the gap.\label{4band}}
\end{figure}

The above major results still hold in the four-orbital model as
summarized in Fig.~\ref{4band}(a). The pairing symmetry is still 
$A_{1g}$ $s$-wave and with a sign change between the $d_{xy}$ and $d_{xz(yz)}$ orbitals. As shown in Fig.~\ref{4band}(a), the positive pairing energy is achieved when $U$ reaches about $0.4W$ as well. The NN-AFM is relatively stronger in the four orbital model than the one in the three-orbital model since the NNN hopping is weaker. The SC pairing in the fourth orbital $d_{x^2-y^2}$ is
negligible, while 
the SC pairing strength in the $d_{xy}$ orbital becomes stronger, compared to the three-orbital model.

To understand the signs of SC order parameters of different orbitals, we can consider the effective Josephson coupling between them. The Josephson couplings can be derived from inter-orbital hoppings.  Let's consider the inter-orbital hopping between arbitrary two sites (denoted by A and B) described by
$H_{h}=t^{AB}_{\alpha\beta}(d^\dag_{A,\beta,\sigma}d_{B,\alpha,\sigma}+H.c)$,
where $\alpha, \beta$ are two different orbitals. Assuming it is
half-filled for both orbitals at the two sites and the spin singlet
is formed within each orbital due to the intra-orbital AFM coupling between two sites, we can easily show the inter-orbital hopping can generate the effective Josephson coupling between two orbitals given by
\be
H^{eff}=-2\frac{t^{AB}_{\alpha\beta}t^{AB}_{\beta\alpha}}{U^\prime}
\Delta_\alpha^\dag\Delta_\beta, \label{effjosephson}
\ee
where $\Delta_{\alpha}$ are the SC order
parameters. In the case of iron-based superconductors, the general
symmetry of the lattice requires 
$t^{ij}_{xz,yz}=t^{ij}_{yz,xz}$ and
$t^{ij}_{xy,xz(yz)}=-t^{ij}_{xz(yz),xy}$, as sketched in Fig.~\ref{checkerboard}. Thus the Josephson coupling between the $xz$ and $yz$ is negative while the one between the $xz(yz)$ and $xy$ is positive, which favors the sign change between $\Delta_{xz(yz),xz(yz)}(ij)$ and $\Delta_{xy,xy}(ij)$. This explains the
numerical results.

{\it Discussion.}- It is very interesting to compare our study to other previous theoretical studies. First, our study confirms that the pairing symmetry is dominated by the $A_{1g}$ $s$-wave induced by the NNN-AFM coupling, which will lead to the sign change between electron and
hole pockets as many previous studies have
concluded\cite{seo-08prl206404,Berg2010,wang-09prl047005,thomaleasvsp}.
Second, our study suggests that the Hund's coupling is also important
in inducing the SC pairing and the NNN-AFM correlation\cite{Haule2009}. Without the Hund's coupling, the model does not lead to the SC instability. Third, our study indicates the importance of the $d_{xy}$ orbital in the strong coupling limit. The pairing in the $d_{xy}$ orbital favors an opposite sign to those in
the $d_{yz}$ and $d_{xz}$ orbitals. This result,  if still holds in the homogeneous limit, may
also cause an anisotropic SC gap in the electron pockets at the $M$ point
of reciprocal space since the electron pockets include both the $d_{xy}$ and $d_{xz(yz)}$ orbitals. The presence of $d_{xy}$ can change many important properties of the SC pairing.
 Indeed, the importance of $d_{xy}$ orbital has been emphasized in the weak coupling approaches, such as the
 fRG which shows that  in the absence of 
the hole Fermi pocket from $d_{xy}$, an anisotropic gap can be developed,  similar to our implication.
However, the physical mechanism, as discussed earlier, is completely different from the one in a weak coupling  approach. 
Finally, 
 it is worth mentioning that the  the signs of the Josephson couplings obtained in eq.\ref{effjosephson}
  explain why the SC state with time reversal symmetry (TRS) breaking as suggested in \cite{Leewc2009,Goswami2010} 
  for iron-SCs is not favored.
    A TRS broken state is favored if and only if the couplings between all three 
orbitals are positive.

Our study provides a new picture of the SC order parameters.  In the context of iron-SCs, the
 $A_{1g}$ $s$-wave pairing is generally agreed to exhibit the sign-change between electron
 and hole Fermi pockets via spin fluctuations.
However, our study,  though from inhomogeneous limit, suggests that in the presence of the $d_{xy}$ orbital, there is an additional sign-change between the $d_{xy}$ and the $d_{xz(yz)}$ orbitals. Considering the recent ARPES experiments, the large hole pocket at the $\Gamma$ point is mainly from $d_{xy}$ orbital while the small two hole pockets which are almost doubly degenerated is composed of the $d_{xz}$ and $d_{yz}$ orbitals\cite{zhangy2009cc}.
 Keeping these orbital characters in mind,
our results do provide a testable, experimental prediction:
the signs  in SC order parameters are {\it opposite} for the large and small hole pockets at the $\Gamma$ point.
We present a schematic plot for SC order parameters in reciprocal space in Fig.~\ref{4band}(b).
More physical consequences associated with such a sign change will be explored in the future study.

We thank A. B. Bernevig,  S. Kivelson,  H. Ding, D.L. Feng, X.H.
Chen and P.C. Dai  for help in discussion.  JPH thanks E. Demler
for asking question related to this study.  YJJ acknowledge the support from NSF, Zhejiang (No.Y7080383) and  NSF, China (No.11004174).


\begin{thebibliography}{10}
%
\bibitem{kamihara-08jacs3296}
Y. Kamihara, {\it et~al}, J. Am. Chem. Soc. {\bf
  130},  3296  (2008).

\bibitem{seo-08prl206404}
K. Seo, B.~A. Bernevig, and J. Hu, Phys. Rev. Lett. {\bf 101},  206404  (2008).

\bibitem{si-08prl076401}
Q. Si and E. Abrahams, Phys. Rev. Lett. {\bf 101},  076401  (2008).

\bibitem{Berg2010}
E. {Berg}, S.~A. {Kivelson}, and D.~J. {Scalapino}, Phys. Rev. B {\bf 81},
  172504  (2010).

\bibitem{Fang2008}
C. Fang {\it et~al.}, Phys. Rev. B {\bf 77},  224509  (2008).

\bibitem{Xu2008a}
C. Xu, M. Mueller, and S. Sachdev, Phys. Rev. B {\bf 78},  02051  (2008).

\bibitem{Haule2009}
K. {Haule} and G. {Kotliar}, New Journal of Physics {\bf 11},  025021  (2009).

\bibitem{Haule2008}
K. {Haule}, J.~H. {Shim}, and G. {Kotliar}, Phys. Rev. Lett {\bf 100},
  226402  (2008).

\bibitem{dai2008j}
J. {Dai}, Q. {Si}, J.-X. {Zhu}, and E. {Abrahams}, PNAS {\bf 106},  4118
  (2009).

\bibitem{mazin-08prl057003}
I.~I. Mazin, {\it et ~al}, Phys. Rev. Lett. {\bf
  101},  057003  (2008).

\bibitem{kuroki-08prl087004}
K. Kuroki {\it et~al.}, Phys. Rev. Lett. {\bf 101},  087004  (2008).

\bibitem{wang-09prl047005}
F. Wang {\it et~al.}, Phys. Rev. Lett. {\bf 102},  1047005  (2009).

\bibitem{thomaleasvsp}
R. Thomale, C. Platt, W. Hanke, and B.~A. Bernevig, arXiv:1002.3599
  (unpublished).

\bibitem{cvetkovic-09epl37002}
V. Cvetkovic and Z. Tesanovic, Europhys. Lett. {\bf 85},  37002  (2009).

\bibitem{korshunov-08prb140509}
M.~M. Korshunov and I. Eremin, Phys. Rev. B {\bf 78},  140509  (2008).

\bibitem{chubukov-08prb134512}
A.~V. Chubukov, D.~V. Efremov, and I. Eremin, Phys. Rev. B {\bf 78},  134512
  (2008).

\bibitem{scalapino1995}
D.~J. Scalapino, Phys. Rep. {\bf 250},  329  (1995).

\bibitem{scalapino1996}
D.~J. Scalapino and S.~A. Trugman, Philos. Mag. B {\bf 74}, 607 (1996).

\bibitem{tsai2006}
W.-F. Tsai and S.~A. Kivelson, Phys. Rev. B {\bf 73},  214510  (2006).

\bibitem{yaoh2007}
H. Yao, W.-F. Tsai, and S.~A. Kivelson, Phys. Rev. B {\bf 76},  161104(R)  (2007).

\bibitem{altman2002}
E. Altman and A. Auerbach, Phys. Rev. B {\bf 65},  104508  (2002).

\bibitem{Cruz2008}
C. de~la Cruz {\it et~al.}, Nature {\bf 453},  899  (2008).

\bibitem{Zhaoj2009}
J. {Zhao} {\it et~al.}, Nature Physics {\bf 5},  55  (2009).

\bibitem{Zhaojun2008}
J. {Zhao} {\it et~al.}, Phys. Rev. Lett. {\bf 101},  167203  (2008).

\bibitem{Ding2008a}
H. Ding {\it et~al.}, Europhs. Lett. {\bf 83},  47001  (2008).

\bibitem{Zhang2010d}
Y. Zhang {\it et~al.}, Phys. Rev. Lett. {\bf {105}},    ({2010}).

\bibitem{Nakayama2010}
K. {Nakayama} {\it et~al.}, arXiv:1009.4236  (2010).

\bibitem{raghu2008}
S. Raghu {\it et~al.}, Phys. Rev. B {\bf 77},  22503  (2008).

\bibitem{daghofer2010}
M. {Daghofer}, {\it et ~al},
 Phys. Rev. B {\bf 81},
  014511  (2010).

\bibitem{Daghofer2008}
M. Daghofer {\it et~al.}, Phys. Rev. Lett {\bf 101},  237004  (2008).

\bibitem{Fisher1988}
D. Fisher and P.~C. Hohenberg, Phys. Rev. B {\bf 37},  493  (1988).

\bibitem{Lee2008a}
P.~A. {Lee} and X.-G. {Wen}, Phys. Rev. B {\bf 78},  144517  (2008).

\bibitem{Leewc2009}
W.~C. Lee, S.~C. Zhang, and C. Wu, Phys. Rev. Lett {\bf 102},  217002  (2009).

\bibitem{Goswami2010}
P. Goswami, P. Nikolic, and Q. Si , Europhys. Lett. {\bf 91},  37006  (2010).

\bibitem{zhangy2009cc}
Y. Zhang {\it et~al.}, arXiv:0904.4022  (2009).

\end{thebibliography}

\end{document}